\documentclass[global,twocolumn]{svjour}
\usepackage{graphicx}

\journalname{Applied Physics A}
\begin{document}

\title{Electronic structure and dynamics of optically excited single-wall carbon nanotubes}

\author{A. Hagen\inst{1}, G. Moos\inst{1}, V.\ Talalaev\inst{2}, J.\ W. Tomm\inst{2} \and  T. Hertel\inst{1,3}
}                     

\institute{Department of Physical Chemistry, Fritz-Haber-Institut
der Max-Planck-Society, Faradayweg 4-6, D-14195 Berlin, Germany
\and Max-Born-Institut f\"{u}r Nichtlineare Optik und
Kurzzeitspektroskopie, Max Born Str. 2 A, D-12489 Berlin, Germany
\and Department of Physics and Astronomy, Vanderbilt University,
Nashville, Tennessee, USA }

\date{Received: date / Revised version: date}

\maketitle

\begin{abstract}
We have studied the electronic structure and charge-carrier
dynamics of individual single-wall carbon nanotubes (SWNTs) and
nanotube ropes using optical and electron-spectroscopic
techniques. The electronic structure of semiconducting SWNTs in
the band-gap region is analyzed using near-infrared absorption
spectroscopy. A semi-empirical expression for $E_{11}^{\rm S}$
transition energies, based on tight-binding calculations is found
to give striking agreement with experimental data. Time-resolved
PL from dispersed SWNT-micelles shows a decay with a time constant
of about 15\,ps. Using time-resolved photoemission we also find
that the electron-phonon ({\it e-ph}) coupling in metallic tubes
is characterized by a very small {\it e-ph} mass-enhancement of
0.0004. Ultrafast electron-electron scattering of photo-excited
carriers in nanotube ropes is finally found to lead to internal
thermalization of the electronic system within about 200\,fs.
\end{abstract}

\section{Introduction}
\label{intro}

The electronic properties of single-wall carbon nanotubes (SWNTs)
and specifically of semiconducting SWNTs hold promise for
application in a variety of optical devices such as
light-emitting- \cite{Misewich03} and photo-sensitive diodes
\cite{Freitag03}, for nonlinear optical- or light harvesting
materials and more. The electronic structure and the dynamics of
optically excited states in SWNTs naturally plays a decisive role
for the operation of such devices. In particular the scattering
times of photoexcited carriers with electrons, phonons and defects
as well as the competition of radiative with non-radiative decay
processes are of great interest for a better understanding of
electronic transport phenomena in optical devices
\cite{Hertel00,Hertel00b,Mele00,Sun00,Hertel02,Vivien02,Jeon03,Kane03,Lauret03,Lebedkin03,Hagen03,Moos03}.

\par Here, we make use of the recent development of techniques for the
exfoliation of SWNTs from nanotube-ropes which provides the basis
for a detailed investigation of optical properties and excitations
in individual SWNTs using near-infrared (NIR) absorption
spectroscopy or spectrofluorimetric techniques
\cite{Connell02,Bachilo02}. These advances also allow to address
questions of fundamental interest such as the character of optical
excitations in systems with reduced dimensionality. Binding
energies of Wannier-type excitons, for example, are predicted to
depend strongly on the dimensionality and the lowest state is
expected to diverge towards infinite binding energy as the system
becomes truly one-dimensional \cite{Loudon59,He91}. It has
previously been speculated that some optical transitions in
individual SWNTs are excitonic, but so far, the reported evidence
is not conclusive. Here, we will discuss the character of optical
excitations in individual SWNTs in further detail. The
corresponding transition energies are analyzed using tight-binding
(TB) calculations which
 show that the observed variation of transition
energies with tube diameter and chirality can be identified with
variations in the TB band-gaps. The decay of optically excited
states is determined by the competition of non-radiative and
radiative decay processes. These are investigated experimentally
using picosecond time-resolved photoluminescence (PL)
spectroscopy.

\par
We have furthermore studied carrier dynamics in SWNT ropes using
femtosecond time-resolved photoemission.  A detailed analysis of
these experiments allows to determine characteristic timescales of
fundamental scattering processes such as electron-electron ({\it
e-e}) or electron-phonon ({\it e-ph}) scattering. We find that
{\it e-e} interactions in bundles of SWNTs give rise to {\it e-e}
scattering times which strongly increase as the carrier energy
approaches the Fermi level. The internal thermalization of the
photo-excited electron gas is facilitated by such {\it e-e}
scattering processes and is characterized by a time constant of
200\,fs. The electron-phonon ({\it e-ph}) coupling in metallic
tubes finally is found to be extraordinarily weak with a mass
enhancement of only 0.0004. The latter suggests that room
temperature {\it e-ph} scattering times are on the order of
15\,ps. These results are discussed with respect to the influence
of radiative and non-radiative decay processes on the carrier
dynamics in individual tubes and tube ropes.

\section{Experimental}
\label{sec:1}
Ultraviolet to visible to near-infrared (UV-VIS-NIR)
absorption spectra and time-resolved PL studies were obtained from
samples synthesized by the high pressure CO decomposition
technique \cite{Nikolaev99}. So called HiPCO material has been
purchased from CNI Houston. For absorption experiments, individual
SWNTs were isolated from HiPCO material in sodiumdodecylsulfate
(SDS) micelles by the technique described in ref.
\cite{Connell02}. At wavelengths below 1350 nm the well resolved
spectrum from micelles in ${\rm H_2O}$ was extended by results
from measurements in ${\rm D_2O}$. For PL measurements the samples
were dispersed using sodium cholate. Time-resolved PL from SWNT
cholate micelles is performed using 300\,mW of output from a
Tsunami Ti:sapphire with sub-100 fs excitation pulses at a
repetition rate of 82 MHz and a wavelength of 795\,nm. Detection
is done by a Hamamatsu synchroscan streak camera with an
IR-enhanced cathode. The time resolution is here found to be
slightly better than 10\,ps. The laser spot size in the 2\,mm long
quartz cuvette was $\approx 0.5\,$mm in diameter. This allowed to
keep pulse fluences at a very low level of $10^{13}$\, photons per
$\rm cm^2$ or less.

\par
Time-resolved photoemission experiments were here performed with
samples consisting of purified nanotube ropes that were produced
by the laser vaporization technique (tubes@rice). A Ti:sapphire
laser pumped optical parametric amplifier (Coherent) generates
tunable femtosecond pulses with wavelengths between 470\,nm and
730\,nm. Fundamental and second harmonic beams generated from the
OPA output are used as visible pump and UV probe respectively. The
fluence for the visible pump and UV probe beams is typically
50\,$\mu$J\,cm$^{-2}$ and 5\,$\mu$J\,cm$^{-2}$, respectively,
which corresponds to a pump power of ~ $10^9 $W\,cm$^{-2}$. For
more details see reference \cite{Hertel02}.

\section{Results and discussion}
\label{sec:2}

\begin{figure}[!]
\includegraphics[width=8cm]{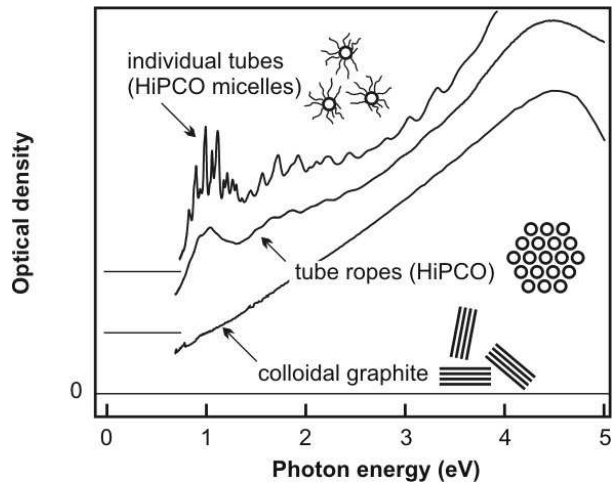}
\caption{Comparison of the optical density of colloidal graphite
(lower trace), crystalline single-wall carbon nanotube ropes
(middle trace) and exfoliated SWNTs in solution.}\label{Fig1}
\end{figure}

In Fig. \ref{Fig1} we compare the absorption spectrum from
colloidal graphite with spectra from SWNT-ropes as well as from
individual exfoliated SWNT-micelles in aqueous solution. The
optical density of colloidal graphite increases continuously from
the near infrared to the UV range due to the logarithmic
singularity in the joint density of states near 5\, eV for
excitation  of free carriers from $\pi$ to $\pi^*$ bands
\cite{Daniels70,Ahuja97}. The optical density in SWNT ropes
follows a similar general trend as seen for graphite illustrating
the intimate relationship of the electronic structure of these two
materials. However, the middle spectrum in Fig. \ref{Fig1} from
SWNT ropes also exhibits some pronounced fine structure on top of
the continuously rising background. These features can be
attributed to optical transitions between different sub-bands of
semiconducting and metallic SWNTs in the suspension. For tubes
with a diameter of approximately 1\,nm, absorption features in the
range from 0.8\,eV to 1.3\,eV and features in-between about
1.4\,eV and 2.5\,eV are associated with $E_{11}^{S}$ and
$E_{22}^{S}$ transitions between the first and second $\pi$
sub-bands in semiconduction tubes, respectively. Broad and not
clearly resolved features between about 2.0\,eV to 3.5\,eV are
 due to $E_{11}^{M}$ transitions between the first sub-bands
in metallic tubes. The nomenclature used here for optical
transitions is $E_{kl}^{X}$, where $k$ and $l$ refer to the
valence and conduction sub-band index, respectively and $X=S,M$
refers to transitions in semiconducting or metallic tubes,
respectively. Transitions with $k=l$ are dipole-allowed for light
polarized parallel to the tube axis \cite{Lin00}.

\par
If individual SWNTs are combined into ropes one finds that
$E_{11}^{S}$ transitions become substantially broadened. The
corresponding 100\,meV increase of the FWHM of absorption features
is indicative for the large contribution of tube-tube interactions
to the width of transitions in spectra from SWNT ropes
\cite{Hagen03}. Higher lying $E_{22}^{S}$ and $E_{11}^{M}$
transitions are broad both in the exfoliated and crystalline phase
which suggests that intratube relaxation from higher energy states
provides an efficient decay channel for tubes in all environments.
In the following we will begin the discussion with a detailed
analysis of $E_{11}^{S}$ transitions within the first absorption
cluster.

\subsection{Analysis of absorption spectra from SWNT-micelles}
\label{sec:3}

\begin{figure}[!]
\includegraphics[width=8cm]{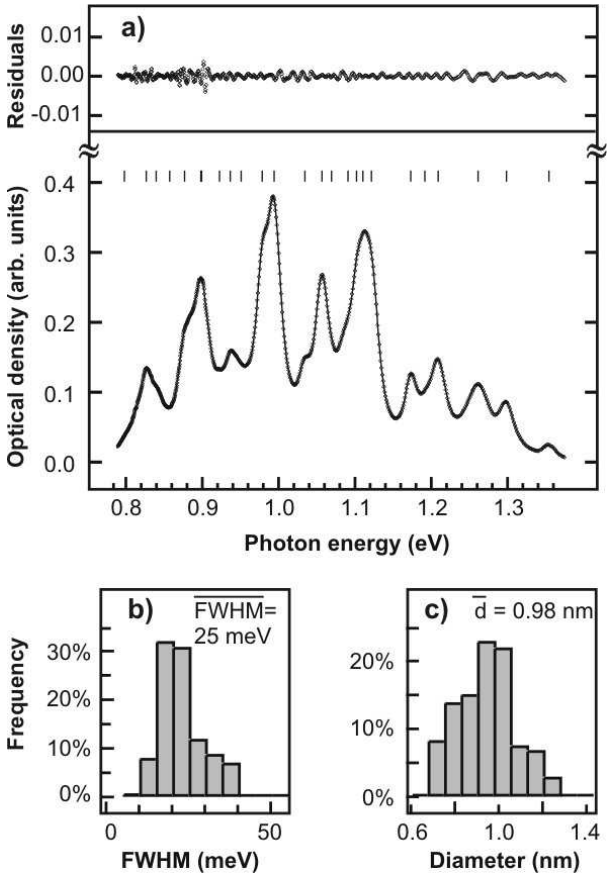}
\caption{a) Near-infrared region of the background corrected SWNT
micelle absorption spectrum (circles) and multi-peak fit using a
linear combination of 24 Voigt profiles (thin line). Individual
peak positions are marked by vertical lines above the spectrum. b)
the resulting diameter- and c) peak-width distributions for
semiconducting SWNTs.}\label{Fig2}
\end{figure}

In Fig. \ref{Fig2} we have plotted a background corrected
absorption spectrum with well-resolved spectral features in the
range of the first absorption cluster. To analyze the optical
transitions in this energy range in further detail we performed a
multi-peak fit with a maximum of 24 independent Voigt profiles.
The resulting fit is shown in Fig. \ref{Fig2}a) as thin line
superimposed on the experimental data (dots). The excellent
agreement between fit and experiment can be appreciated from the
nearly vanishing residuals in the upper part of Fig. \ref{Fig2}a).
Due to the excellent signal to noise ratio, errors of peak
positions are on average only about 1\,meV. The experimental peak
energies obtained here can be found in Table \ref{tab:1} together
with transitions obtained from PL measurements \cite{Bachilo02}.
The distribution of peak widths with its mean at 25\,meV is shown
in Fig. \ref{Fig2}b). The contribution of Gaussian and Lorentzian
components to the Voigt profiles is about equal. Similar average
widths of 23\,meV and 25\,meV were recently also observed for
band-gap PL from single SWNTs dispersed on a surface or SWNT
micelle ensembles in solution, respectively
\cite{Bachilo02,Hartschuh03}.

\begin{table}
\caption{Comparison of the position of spectral absorption
features obtained experimentally and from the semi-empirical, TB
band-structure based model.}
\begin{tabular}{llll}
\hline\noalign{\smallskip}
Tube type &  $d ~({\rm nm})$  & $h\nu_{11}^{expt}$ (eV)  & $h\nu_{11}^{calc}$ (eV)  \\
\noalign{\smallskip}\hline\noalign{\smallskip}
(5,4)   &   0.62  &   $1.485 \footnotemark[1]     $&   1.480  \\
(6,4)   &   0.69   &   $1.417 \footnotemark[1]    $&   1.421  \\
(9,1)   &   0.76   &   $1.354\pm0.001 $  &   1.357  \\
(8,3)   &   0.78   &   $1.298\pm0.001 $  &   1.302  \\
(6,5)   &   0.76   &   $1.261\pm0.001 $  &   1.260  \\
(7,5)   &   0.83   &   $1.208\pm0.001 $  &   1.208  \\
(11,0)   &   0.87   &  $ 1.190\pm0.001$   &   1.187  \\
(10,2)   &   0.88   &  $ 1.172\pm0.001$   &   1.170  \\
(9,4)   &   0.91   &   $1.120\pm0.007 $  &   1.123  \\
(8,4)   &   0.84   &   $1.109\pm0.002$   &   1.108  \\
(7,6)   &   0.89   &   $1.100\pm0.001$   &   1.097  \\
(9,2)   &   0.81   &   $1.089\pm0.008$   &   1.095  \\
(12,1)   &   0.99   &  $ 1.067\pm0.001$   &   1.053  \\
(8,6)   &   0.97   &   $1.054\pm0.001 $  &   1.053  \\
(11,3)   &   1.01   & $  1.032\pm0.001$   &   1.032  \\
(10,3)   &   0.94   &  $ 0.989 \footnotemark[1]     $ &   0.985  \\
(10,5)   &   1.05   & $  0.989 \footnotemark[1]    $  &   0.988  \\
(9,5)   &  0.97   &   $0.991\pm0.001  $ &   0.988  \\
(11,1)   &   0.91   &  $ 0.979 \footnotemark[1]     $ &   0.976  \\
(8,7)   &   1.03   &  $ 0.976\pm0.001 $  &   0.971  \\
(13,2)   &   1.12   & $  0.948\pm0.003$   &   0.947  \\
(9,7)   &   1.10   &  $ 0.933\pm0.001 $  &   0.935  \\
(12,4)   &   1.14   & $  0.919\pm0.001$   &   0.924  \\
(10,6)   &   1.11   & $  0.895 \footnotemark[1]    $  &   0.889  \\
(11,4)   &   1.07   & $  0.896\pm0.001$   &   0.897  \\
(12,2)   &   1.04   & $  0.895\pm0.002$   &   0.896  \\
(11,6)   &   1.18   & $  0.884 \footnotemark[1]    $  &   0.885  \\
(9,8)   &   1.17   &  $ 0.873\pm0.001 $  &   0.873  \\
(15,1)   &   1.23   & $  0.867 \footnotemark[1]    $  &   0.871  \\
(13,5)   &   1.28   & $  0.835 \footnotemark[1]    $  &   0.836  \\
(10,8)   &   1.24   & $  0.836\pm0.007$   &   0.842  \\
(12,5)   &   1.20   & $  0.826 \footnotemark[1]    $  &   0.820  \\
(13,3)   &   1.17   & $  0.823\pm0.001$   &   0.826  \\
(10,9)   &   1.31   & $  0.794\pm0.004$   &   0.792  \\
\noalign{\smallskip}\hline
\end{tabular}
\label{tab:1}
\end{table}
\footnotetext[1] {Values taken from ref. \cite{Connell02} are
redshifted by 3\,meV.}

\par
A comparison of peak positions obtained here with those reported
by Bachilo et al. for PL measurements from SWNT-micelles dispersed
in D$_2$O \cite{Bachilo02} yields extraordinary agreement if a
~3\,meV blue-shift of the fluorimetric data is accounted for. The
previously observed redshift of luminescence data of a few meV
\cite{Connell02} here appears to be overcompensated by differences
in transition energies in different environments, i.e. in heavy
and light water. The peak assignment adopted here is practically
identical to that previously discussed in ref. \cite{Hagen03} as
'scenario I' and is guided by the energetic order of peaks in
fluorimetric data of Bachilo et al. \cite{Bachilo02}. In cases
where no peaks are reported from PL data, our assignment was
guided by the order of peaks expected from the semi-empirical TB
band-structure model discussed below. If the spectral weight of
individual peaks in the absorption spectrum is combined with the
assignment to specific tube types, i.e. tube diameters, we can
also calculate the diameter distribution of semiconducting tubes
in the present sample which is reproduced in Fig. \ref{Fig2}c).
The mean diameter of 0.98\,nm obtained here is in good agreement
with that recently obtained by a luminescence study 0.96\,nm
\cite{Bachilo02} and Raman studies, with 1.13\,nm \cite{Gregan02}
and 0.98\,nm \cite{Kukovecz02}. The diameter distribution obtained
in ref. \cite{Hagen03} is at variance with that obtained here,
apparently because the statistical weight of individual tube types
varies if the JDOS --- as in a previous study \cite{Hagen03} ---
or Voigt profiles are used to fit experimental spectra. The
probability distribution of the chiral vector map resulting from
the above assignment and fit shows a slight preference for
armchair tubes.

\par
Next, we will attempt a quantitative discussion of observed
transition energies by comparison with energies predicted by a
refined version \cite{Ding02,Hagen03} of the original TB
band-structure model \cite{Saito92,Hamada92}. The TB calculations
used here explicitly include the crucial diameter and chirality
dependence of the TB hopping integral $\gamma_0$ without which any
such comparison would be useless. The refined TB-model leads to
more pronounced oscillations of band-gaps with tube diameter than
predicted by the earlier zone folding schemes because the latter
used to neglect mutual realignment of $p_z$ orbitals on the curved
graphene sheet. However, the magnitude of $\gamma_0^{\infty}$, the
TB hopping integral for the uncurved graphene sheet, is frequently
used as a free parameter and there is no general agreement on the
appropriate value that should be used for optical excitations.
Scanning tunneling spectroscopy of individual SWNTs yields
$\gamma_0^{\infty}$ values between 2.45\,eV and 2.7\,eV
\cite{Odom98,Wildoer98}, while an analysis of optical experiments
suggest that values between 2.65\,eV and 3.0\,eV
 are more appropriate (see, for example, refs. \cite{Hertel02} and \cite{Liu02}). Resonant Raman spectra are
frequently analyzed using a hopping integral around 3.0\,eV
\cite{Dresselhaus99}.

\begin{figure}[!]
\includegraphics[width=8cm]{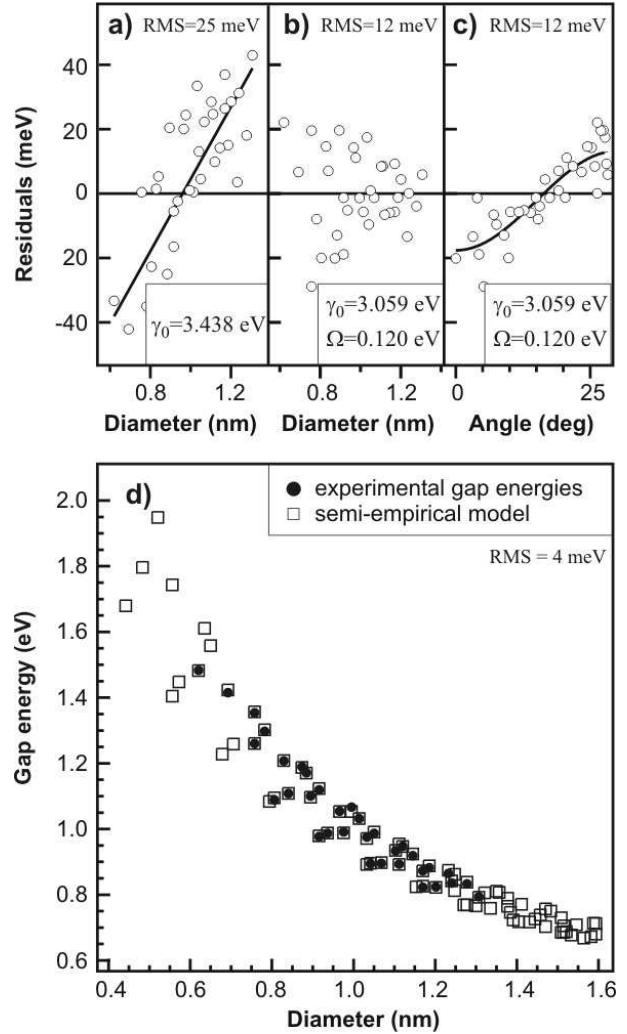}
\caption{a-c) Residuals from a fit of experimental data  with
transition energies from a semi-empirical TB band-structure model.
The agreement between experiment and model becomes increasingly
better if a phenomenological band offset parameter as well as a
chirality correction are included. The straight line in a) is the
linear regression and the solid curve in c) represents the
contribution from the phenomenological chirality correction.
Experimental transition energies are compared in d) with
predictions from the four parameter semi-empirical model. The root
mean square deviation (RMS) of experimental and model transition
energies is also indicated.}\label{Fig3}
\end{figure}

\par
If experimental peak positions are here fit by the one-parameter
TB model one obtains best agreement with calculated gaps if
$\gamma_0^{\infty}=3.450$\,eV. The resulting residuals can be
characterized by a 25\,meV root mean square (RMS) and are plotted
as a function of tube diameter in Fig. \ref{Fig3}a). The steep
slope of the straight line fit to residuals clearly illustrates
that the one-parameter TB model cannot account for the observed
diameter dependence of transition energies. This deficiency can be
removed by the addition of a number of empirical corrective terms
which give successively better agreement with experiment and
thereby also allow to discuss the possible physical origin of
remaining discrepancies. Previously, a first order correction was
introduced in the form of an empirical offset parameter $\Omega$
\cite{Hagen03} which, in combination with a variable
$\gamma_0^{\infty}$, allow to skew the diameter dependence of
calculated transition energies (see Fig. \ref{Fig3}b)). Within
this two parameter model, best agreement is obtained for
$\gamma_0^{\infty}=3.059\,$eV and $\Omega=0.120\,$eV. However, the
resulting residuals exhibit a clear correlation with the tube
chiral angle $\theta(n,m)=\arcsin(\sqrt{3}\,m
/(2\cdot\sqrt{n^2+n\cdot m+m^2}))$ as seen in Fig. \ref{Fig3}c).
The second order correction thus includes a term which varies with
the chiral angle as $A \cdot \cos(6\cdot\theta(n,m))$ (see solid
line in Fig. \ref{Fig3}c). At this point the root mean square
deviation of experimental and calculated transition energies is
only 6\,meV ($\gamma_0^{\infty}=3.061\,$eV, $\Omega=0.118\,$eV,
$A=-0.0154\,$eV). The last corrective term $B\cdot d(n,m) \cdot
\cos(6\cdot\theta(n,m))$ is included to partially remove a
combined correlation of residuals with diameter $d$ and chiral
angle $\theta$. The phenomenological four parameter model
describing optical transition energies $E_{11}^{S}$ in the
band-gap region is finally given by:

\begin{eqnarray}
E_{11}^{\rm opt\,S}(n,m)&=&\gamma_0^{\infty} \cdot
\epsilon_{11}^{\rm TB}(n,m)
\\
&+& (A+B\cdot d(n,m))\cos[6\cdot\theta(n,m)] \nonumber
\\
&+& \Omega  \nonumber \label{Gapform}
\end{eqnarray}

\noindent where $\epsilon_{11}^{\rm TB}$ are the TB band-gaps as
calculated using the diameter and chirality corrected nearest
neighbor hopping integrals \cite{Ding02}. Best agreement with
experiment is here obtained for $\gamma_0^{\infty}=3.022\,$eV,
$\Omega=0.130\,$eV, $A=-0.0543\,$eV, and $B=0.00395\,$eV. The
quality of the agreement between experiment and calculated
transition energies is evident from Table \ref{tab:1} and Fig.
\ref{Fig3}d) where calculated gap-energies are plotted as a
function of the tube diameter. Note, that the root mean square
deviation is below 4\,meV.

\begin{figure}[!]
\includegraphics[width=8cm]{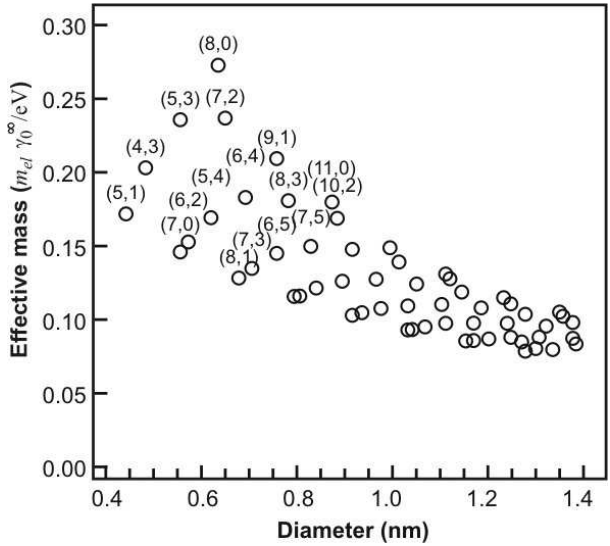}
\caption{Effective exciton masses calculated using the chirality
and diameter corrected TB-band-structure model \cite{Ding02}. In
contrast to expectations for excitonic transitions, there is no
apparent correlation of transition energies with effective
masses.}\label{Fig4}
\end{figure}

\par
Alternatively, we have also combined diameter and chirality
corrected TB calculations with a simple Wannier type exciton
model. Transition energies are here calculated within the
effective mass approximation allowing also for fractal
dimensionality of excitons $\alpha$, with $1<\alpha<3$. The
transition energy is then given by:

\begin{equation}
E_{11}^{S}(n,m)=\gamma_0^{\infty} \cdot \epsilon_{11}^{\rm
TB}(n,m) - \frac{\mu\,E_H}
{\epsilon^2\,m_{el}\left(n+\frac{\alpha-3}{2}\right)^2}
\label{Excitonform}
\end{equation}

\noindent where $n=1,2,3,...$ is the principal exciton quantum
number while $\mu$ and $m_{el}$ are the effective exciton and free
electron mass and $\epsilon$ is the dielectric constant
\cite{He91}. $E_H$ is the Rydberg constant. $\mu(n,m)$ is obtained
from TB-valence and conduction band effective masses $m_h$ and
$m_e$ using $\mu^{-1}=m_e^{-1}+m_h^{-1}$. For semiconducting tubes
studied here $\mu$ is found to vary strongly with chirality and
diameter and decreases from values around
$0.25\,m_{el}\,\gamma_0^{\infty}/{\rm eV}$ at 0.6\,nm diameter to
about $0.10\,m_{el}\,\gamma_0^{\infty}/{\rm eV}$ at 1.3\,nm
diameter (see Fig. \ref{Fig4}). The dielectric constant is assumed
to be less dependent on chirality and diameter and the exciton
binding energy should thus be most strongly influenced by its
effective mass. Interestingly, a fit of experimental data to eq.
\ref{Excitonform}, where $\epsilon$ and $\alpha$ are treated as
adjustable parameters provides no significant improvement over the
zero-order one-parameter TB model described above. We also find no
correlation of any of the residuals described above with
$\mu(n,m)$.

\par
This analysis suggests that optical transitions in SWNT micelles
are not excitons of the Wannier type. However, other observations
are in favor of such an interpretation as for example the large
value of $\gamma_0^{\infty}=3.438\,$eV in the one-parameter fit.
It can hardly be reconciled with previously reported values. In
particular band-gaps observed in scanning tunneling spectroscopy
are at variance with the transition energies measured here. The
band-gap measured by STS is determined by the threshold for
injection of free electrons or holes into the band-structure of
SWNTs and Coulomb interactions play a comparatively negligible
role. However, Coulomb interactions are crucial for exciton
formation and lead to a renormalization of the band-structure
obtained from non-interacting electron models such as the TB
calculation. The general trend of Coulomb interactions would be to
significantly raise the energy needed for excitation of free
electron-hole ({\it e-h}) pairs. The discrepancy between
transition energies observed in optical and STS experiments would
thus be easily resolved if optical transitions in SWNTs were
excitonic in character. A comparison of the high transition
energies observed here for a specific tube type with results from
STS experiments, suggests that Coulomb interactions lead to a
substantial renormalization of the non-interacting electron
band-structure and gap energies perhaps by as much as ten to tens
of percent. Similar changes of band-gap energies as a consequence
of Coulomb interactions have been predicted theoretically by Ando
\cite{Ando97}.

\par
Here, a further point of interest is the observation that the
absorption spectrum of Fig. \ref{Fig3}a) can be extremely well
described by a linear combination of nearly perfectly symmetrical
absorption features. Note, that absorption by generation of free
electron-hole pairs would be associated with a pronounced
asymmetry of absorption features. Specifically, the joint density
of states (JDOS) shows such asymmetries and consequently also the
absorption cross sections --- even if spectral broadening is
included \cite{Hagen03,Lin00}. The highly symmetrical shape of
absorption features thus appears to be consistent with excitonic
transitions.

\par
In conclusion, we observe striking agreement of experimental
transition energies and energies obtained from a semi-empirical 4
parameter model based on refined TB band-structure calculations.
This strongly suggests that the observed oscillation of transition
energies with diameter and chirality is dictated by corresponding
oscillations of the TB band-gaps. The somewhat unexpectedly large
band-gaps and the highly symmetrical shape of absorption features
in NIR spectra are consistent with an excitonic character of
optical transitions. We speculate, that the offset parameter
$\Omega$ provides a rough measure of the difference between
band-gap widening due to Coulomb interactions and the exciton
binding energy. This would also suggest that exciton binding
energies are only weakly dependent on tube diameter and chirality,
i.e. their effective mass, which appears to be at variance with
expectations for Wannier type excitons. The nature of optical
transitions in the band-gap region thus remains somewhat unclear
and further experimental work seems necessary to unambiguously
identify the character of optical excitations in SWNTs.

\subsection{Charge-carrier dynamics in individual SWNTs}
\label{sec:4}

In the following we will discuss scattering and relaxation
processes of optically excited carriers in individual SWNTs and
SWNT ropes as studied by time-resolved photoluminescence (PL) and
photoelectron (PE) spectroscopy.

\begin{figure}[!]
\includegraphics[width=8cm]{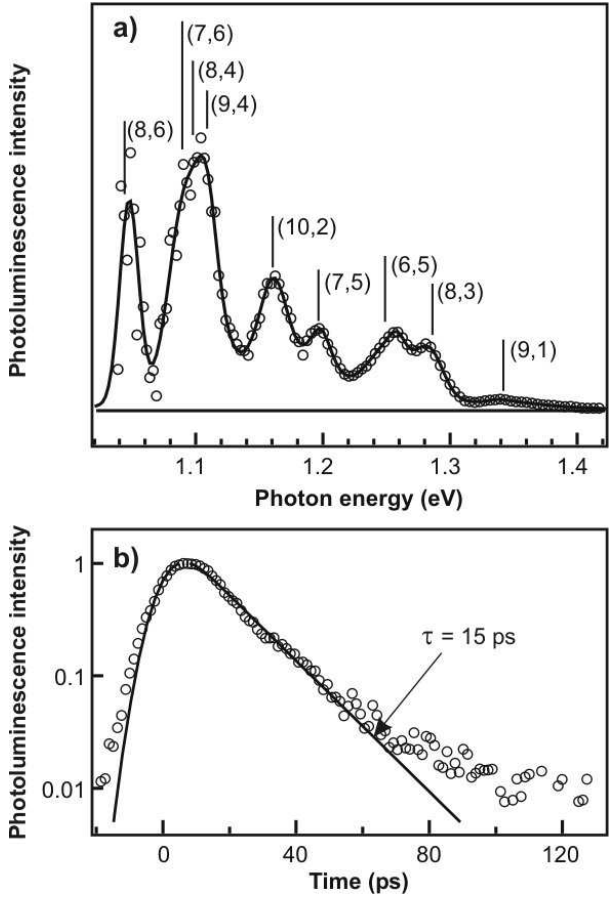}
\caption{a) Photoluminescence spectrum from individual
SWNT-micelles. The name and diameter of a number of tubes is also
specified. b) Energy integrated photoluminescence decay. The decay
has been fit by a 15\,ps decay (solid line).}\label{Fig5}
\end{figure}

\par
If SWNTs are exfoliated from ropes and dispersed in aqueous
solution one finds that the PL quantum yield increases by orders
of magnitude reaching values on the order of
$10^{-3}$.\cite{Connell02} The associated PL spectra are
characterized by a very small Stokes shift of about 5\,meV. In
Fig. \ref{Fig5} a) we have plotted a PL spectrum from water
suspended HiPCO micelles in the wavelength range from 850\,nm to
1200\,nm. The pump power in these experiments corresponds to a
pulse fluence of below $\cdot10^{13}$ photons per cm$^2$ such that
nonlinear effects are not expected. The major features can be
linked to absorption features in the same energy range
corresponding to transitions in tubes with small diameters between
9.1\,\AA and 7.6 \AA. We have marked the peak positions and
diameters of a few tubes in this energy range. The
energy-integrated PL signal in Fig. \ref{Fig5}b) decays with a
time-constant of about 15 ps. Deviations from a simple exponential
decay may be indicative of a distribution of relaxation times in
these samples, i.e. heterogeneous broadening of the corresponding
transitions. Such heterogeneities may arise, for example, from a
distribution of tube lengths. In combination with the recently
reported PL quantum yield of $10^{-3}$ \cite{Connell02}. This
indicates that the radiative lifetime of the luminescent states is
on the order of 10\,ns. The fast decay observed here thus
underlines the importance of non-radiative decay processes for the
PL lifetime. The detailed mechanism of such non-radiative decay
processes, however, needs to be explored in further detail.

\subsection{Charge-carrier dynamics in SWNT ropes}
\label{sec:5}

\begin{figure}[!]
\includegraphics[width=8cm]{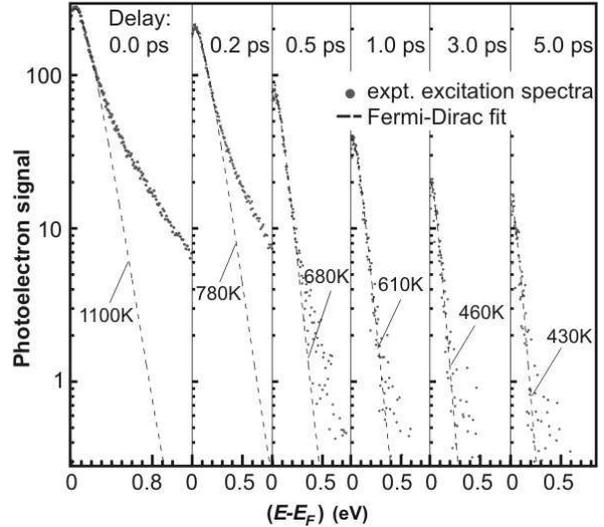}
\caption{Femtosecond time-resolved photoelectron spectra at
different time-delays up to 5\,ps. The contribution of
photoexcited non-thermal electrons is evident from the apparent
deviation of the experimental data from the Fermi-Dirac fit
(dashed line), in particular at early delay times up to about
0.5\,ps}\label{Fig6}
\end{figure}

\par
As mentioned above, the luminescence quantum yield is reduced
severely if tubes are agglomerated in tube ropes were tube-tube
interactions lead to a rapid decay of excited states. This is also
supported by the $\approx 100\,$meV increase of linewidths in
absorption spectra from crystalline SWNT samples \cite{Hagen03}.
The fast decay of excited states can be clearly seen in
femtosecond time-resolved photoelectron spectra as the ones shown
in Fig. \ref{Fig6} the photoelectron signal from the
vicinity of the Fermi level of SWNT ropes is shown on a
logarithmic scale for different pump-probe delays. Here, the
system was perturbed by a visible 2.32\,eV pump pulse and was
probed by a weaker second harmonic 4.64\,eV UV probe pulse. The
photoelectron spectra clearly show the dynamics of the optically
excited SWNT ropes on the sub- and picosecond timescales. The
absence of pronounced features in these spectra similar to those
found in optical spectra can be attributed to the greater
sensitivity of photoemission to the alignment of the
band-structure with respect to the Fermi level
\cite{Hertel00b,Hertel02}.

\begin{figure}[!]
\includegraphics[width=8cm]{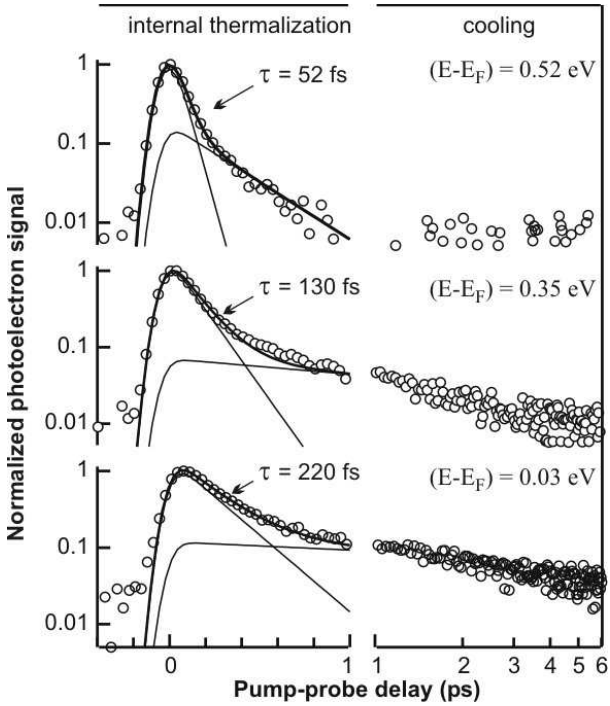}
\caption{Photoelectron cross-correlation traces from SWNT ropes
for three different electron energies. The initial fast component
is seen to be strongly energy dependent and can be associated with
electron-electron scattering leading to rapid internal
thermalization of the laser excited system. The slow component is
associated with electron-phonon interactions and leads to a
somewhat slower cooling of the system back to the lattice
temperature of 300\,K.}\label{Fig7}
\end{figure}

\par If spectra are fit by a hot Fermi-Dirac distribution (dashed
lines) one finds that the electron distribution has a non-thermal
component up to a pump-probe delay of about 0.5\,ps. Here, the
process of internal thermalization of the electronic system after
photoexcitation can be characterized by a time-constant of
200\,fs. We have previously shown that the latter can be
attributed to rapid electron-electron scattering processes similar
to those in graphite \cite{Hertel00b,Hertel02}. At the same time
one also finds that after a rapid initial jump to slightly above
2000\,K the temperature of the electronic system cools back down
towards the lattice temperature of 300\,K on the picosecond
time-scale. This process is intimately related to the
electron-lattice interaction and can be used to quantitatively
determine the electron-phonon mass enhancement parameter
\cite{Hertel00}.

\par
The dynamics of internal thermalization and cooling can also
clearly be discriminated using the time-dependence of the
intensity at a specific photoelectron energy (see Fig.
\ref{Fig7}). Each of the cross-correlation traces in Fig.
\ref{Fig7} monitors the intensity within a certain energy window
in the PE-spectrum. Using the probe photon energy the latter can
be related to a specific intermediate state energy with respect to
the Fermi level. The cross-correlation traces in Fig. \ref{Fig7}
can be decomposed into two components, one decaying on the
subpicosecond and one on the picosecond time-scale. The fast
component is associated with the internal thermalization of the
electronic system while the slow component is related to the
cooling of the laser heated system \cite{Hertel02}.

\begin{figure}[!]
\includegraphics[width=8cm]{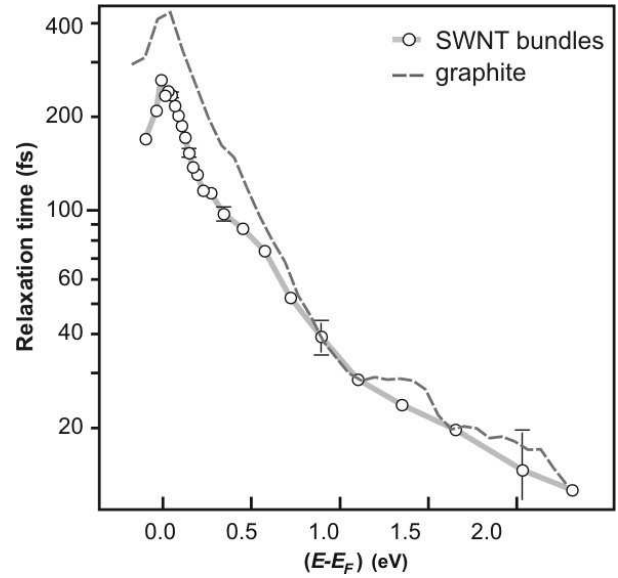}
\caption{Energy dependence of the fast component in
cross-correlation traces from SWNT ropes. The pronounced increase
of relaxation times near the Fermi-level and the resemblance with
results in graphite suggest that the fast component is associated
with electron-electron scattering processes.}\label{Fig8}
\end{figure}

\par
The energy dependence of the fast component is plotted in Fig.
\ref{Fig8} showing a pronounced increase of relaxation-times in
the vicinity of the Fermi-level, a behavior typical of metallic
systems \cite{Hertel96}. Similar decay times are also found in
graphite \cite{Moos01} (see Fig. \ref{Fig8}) where the fast decay
could be attributed to intralayer electron-electron scattering by
a comparison with recent ab inito self-energy calculations
\cite{Moos01,Spataru01}. The fast sub-picosecond decay found here
can readily account for the vanishingly small PL quantum yield
observed for SWNT ropes. The decay of excited states in
semiconducting tubes is believed to be facilitated by rapid
intertube electron scattering and successive generation of low
energy secondary electron cascades, primarily in metallic tubes.
The excess energy of the resulting hot electron distribution is
finally distributed among lattice and electronic degrees of
freedom by the {\it e-ph} interactions discussed in the following
parahraphs.

\par
The slow decay seen in the cross-correlation traces of Fig.
\ref{Fig7} is analyzed in more detail following the approach
described in ref. \cite{Moos03}. To illustrate this in detail, we
have plotted the electronic temperature $T_{e}$ and the
corresponding internal energy of the electronic system as a
function of time before and after laser excitation in Fig.
\ref{Fig9}. The internal energy of the electronic system is seen
to jump to about $10^7\,{\rm W/m^3}$ as a consequence of laser
excitation. If the evolution of the internal energy is
differentiated with respect to time one obtains the rate of energy
change of the electronic system $H=H(T_e,T_l)$, or more
specifically the rate at which energy is transferred between
electrons at temperature $T_e$ and the lattice at temperature
$T_l$ (see inset of Fig. \ref{Fig9}). This rate can be related to
the electron-phonon mass-enhancement parameter $\lambda$ using:

\begin{figure}[!]
\includegraphics[width=8cm]{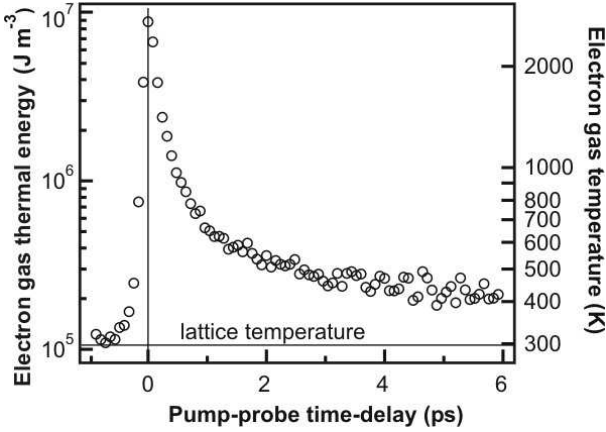}
\caption{Evolution of the internal energy and the corresponding
electron gas temperature of the electronic system after laser
heating.}\label{Fig9}
\end{figure}

\begin{equation}
H(T_e,T_l)=\frac{144\,\zeta(5)\,k_B\,\gamma}{\pi\,\hbar}~\frac{\lambda}{\Theta_D^2}~(T_e^5-T_l^5)\label{Hform}
\end{equation}

\noindent where $\zeta(5)=1.0369...$ is Riemann's Zeta function,
$\gamma$ is the specific electronic heat capacity and $\Theta_D$
the samples Debye temperature \cite{Moos03,Allen87}. Note, that we
here measure the electron-phonon energy transfer in metallic tubes
because metallic tubes contribute most strongly to the density of
states near the Fermi level. For the following analysis we used a
Debye temperature of 1000\,K and a specific heat capacity of
nanotube ropes with a 2:1 mixture of semiconducting to metallic
tubes of $12\,\mu$J/mole/K. A fit to the temperature dependence of
the experimental energy transfer rate using the above constants
yields an extraordinary small $\lambda$ of $0.0004\pm0.0001$ (see
Fig. \ref{Fig9}). This weak coupling is consistent with electron
scattering from to two different acoustic phonon modes: a
longitudinal- and a so called twiston mode \cite{Hertel00}. Note
that the electron-phonon mass enhancement in copper is almost 3
orders of magnitude higher \cite{Brorson90}.

\par
The above value for $\lambda$ can also be used to calculate the
ballistic {\it e-ph} mean free path from the {\it e-ph} scattering
time $\tau_{e-ph}$ which is given by:

\begin{equation}
\frac{1}{\tau_{e-ph}}=24\,\pi\,\zeta(3)\frac{k_B}{\hbar}~\frac{\lambda}{1+\lambda}~\frac{T_e^3}{\Theta_D^2}\label{tauform}
\end{equation}

\noindent With $\zeta=1.2020...$ and the above mass enhancement
parameter this would give a room temperature scattering time in
metallic nanotubes of 15\,ps which corresponds to an {\it e-ph}
mean free path $l_{e-ph}$ of 15\,$\mu$m \cite{Allen75}. This
provides a clear confirmation of previously reported long {\it
e-ph} ballistic mean free paths \cite{Frank98,Appenzeller01} from
the perspective of a time-domain experiment.

\begin{figure}[!]
\includegraphics[width=8cm]{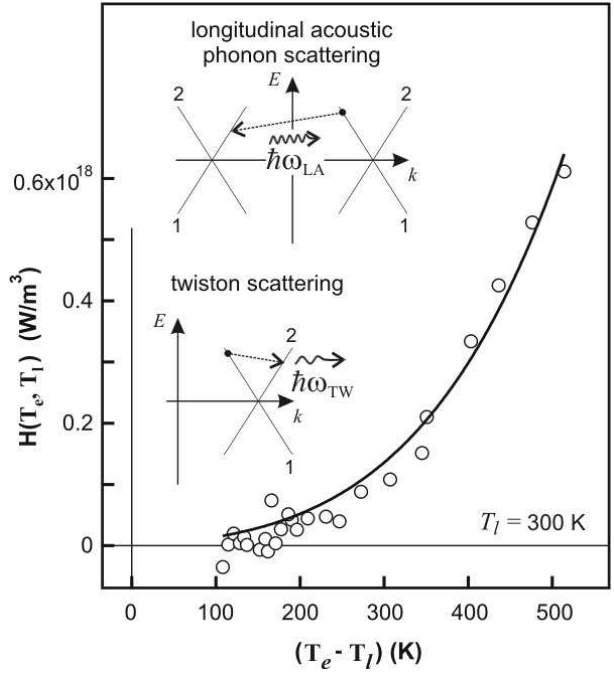}
\caption{Dependence of the electron-to-lattice energy transfer
rate of metallic tubes in tube ropes on the electron gas
temperature. The lattice temperature in this experiment was
300\,K. The solid line is a fit to eq. \ref{Hform} used to
determine the strength of the electron-phonon coupling constant
$\lambda$. Also included is a schematic representation of two
conceivable scattering processes with twistons or longitudinal
acoustic phonons.}\label{Fig10}
\end{figure}

\section{Conclusions}
\label{sec:6}

We have studied the electronic structure and dynamics of optically
excited SWNTs in crystalline ropes as well as isolated in
solution. The transitions in optical spectra of individual SWNTs
can be well described by a semi-empirical, tight-binding based
model with four free parameters. Our analysis of optical spectra
shows that the variation of transition energies with tube diameter
and chirality can be attributed to oscillations of the band-gap
obtained from a refined tight-binding calculation. Note, that
fully empirical expressions of transition energies of the type
given in ref. \cite{Bachilo02} are very useful for the inter- and
extrapolation of band-gaps to presently unobserved tube species
but provide little insight into the physical origin underlying
band-gap variations if different tube types are compared. One of
the corrective terms in the semi-empirical expression presented
here, for example, clearly identifies a chirality dependent
contribution to transition energies whose origin yet needs to be
clarified. Furthermore, in spite of the evidence in favor of an
excitonic character of optically excited states we find no
correlation of the observed transition energies with effective
electron or hole masses as expected for Wannier-Mott type
excitons.

\par The dynamics and energy relaxation in optically excited tube ropes is
found to be dominated by rapid intra- and intertube scattering
processes that lead to internal thermalization of the laser-heated
electronic system within about 200\,fs, most likely through
electron-electron interactions. The coupling of the hot electron
gas to phonons via scattering from longitudinal acoustic phonons
or twistons is found to be characterized by a room temperature
electron-phonon scattering time on the order of 15\,ps. The latter
is a consequence of the weak electron-lattice coupling also
reflected by the experimentally determined {\it e-ph} mass
enhancement parameter $\lambda$ of 0.0004. Relaxation of
photo-excited states near the band-gap of exfoliated
semiconducting tubes is significantly slower than in ropes as
reflected by the commonly observed increase of photoluminescence
quantum yields in isolated tubes to values on the order of
$10^{-3}$. Here, we found the luminescence decay of SWNTs in
sodium cholate micelles to be about 15\,ps which clearly
illustrates that non-radiative decay processes dominate energy
relaxation in semiconducting SWNTs. This also allows to estimate
the radiative lifetime in semiconducting SWNTs to be on the order
of 10\,ns.

\section{Acknowledgements}
It is our pleasure to acknowledge continuing support by G. Ertl.
We also would like to thank G. Dukovic and L. Brus for sharing
experimental absorption spectra prior to publication. We also
thank M. Antonietti as well as O. Bogen and H. Otto for providing
infrastructure for processing nanotube dispersions and I. Dorbandt
for making spectroscopic facilities available to us.


\begin{thebibliography}{}

\bibitem{Misewich03} J. A. Misewich, R. Martel, Ph. Avouris, J. C. Tsang, S. Heinze, and J.
Tersoff, Science {\bf 300}, (2003) 783.

\bibitem{Freitag03} M. Freitag, Y. Martin, J. A. Misewich, R. Martel, Ph.
Avouris, Nano Letters {\bf 3}, (2003) 1067.

\bibitem{Hertel00} T. Hertel and \ G. Moos, \ Phys. Rev. Lett. \ {\bf 84}, \ (2000)\  5002-5005.

\bibitem{Hertel00b} T. Hertel and \ G. Moos, \ Chem. Phys. Lett. \ {\bf 320}, \ (2000)\  359-364.

\bibitem{Mele00} E. J. Mele, P. Kral, and D. Tomanek, Phys. Rev. B {\bf 61},
(2000) 7669.

\bibitem{Sun00} X. Sun, \ Y. N. Xiong, \ P. Chen, \ J. Y. Lin, \ W. Ji, \ J. H. Lim, \ S. S. Yang, \ D. J. Hagan and \ E. W. Van Stryland, \ Appl. Optics \ {\bf 39}, \ (2000)\  1998-2001.

\bibitem{Hertel02} T. Hertel, \ R. Fasel and \ G. Moos, \ Appl. Phys. A-Mater. Sci. Process. \ {\bf 75}, \ (2002)\  449-465.

\bibitem{Vivien02} L. Vivien, \ P. Lancon, \ D. Riehl, \ F. Hache and \ E. Anglaret, \ Carbon \ {\bf 40}, \ (2002)\  1789-1797.

\bibitem{Jeon03} T. I. Jeon, \ K. J. Kim, \ C. Kang, \ S. J. Oh, \ J. H. Son, \ K. H. An,
\ D. J. Bae and \ Y. H. Lee, \ Synth. Met. \ {\bf 135}, \ (2003)\
429-430.

\bibitem{Kane03} C. L. Kane, E. J. Mele, Phys. Rev. Lett. {\bf
90}, (2003) 207401.

\bibitem{Lauret03} J. S. Lauret, \ C. Voisin, \ G. Cassabois, \ C. Delalande, \ P. Roussignol, \ O. Jost and \ L. Capes, \ Phys. Rev. Lett. \ {\bf 90}, \ (2003)\ .

\bibitem{Lebedkin03} S. Lebedkin, F. Hennrich, T. Skipa, and
M. M. Kappes, J. Phys. Chem. B {\bf 107}, (2003) 1949.

\bibitem{Hagen03} A. Hagen and T. Hertel, Nano Lett. {\bf 3}, 2003
383.

\bibitem{Moos03} G. Moos, \ R. Fasel and \ T. Hertel, \ J.
Nanosci. Nanotechnol. \ {\bf 3}, \ (2003)\  145-149.

\bibitem{Connell02} M. \ J. \ O'Connell, S. \ M. \ Bachilo, C. \ B. \ Huffman, V. \ C.\ Moore, M. \ S. \ Strano,
E. \ H. \ Haroz, K. \ L. \ Rialon, P. \ J. \ Boul, W. \ H. \ Noon,
C. \ Kittrell, J. \  Ma, R. \ H. \ Hauge, R. \ B. \ Weisman, R. \
E. \  Smalley, Science, {\bf 279}, 593, (2002).

\bibitem{Bachilo02} S. M. Bachilo, \ M. S. Strano, \ C. Kittrell, \ R. H. Hauge, \ R. E. Smalley and \ R. B. Weisman, \ Science \ {\bf 298}, \ (2002)\  2361-2366.

\bibitem{Loudon59} R. Loudon, Am. J. Phys. {\bf 27}, (1959) 649.

\bibitem{He91} X.-F. He, \ Phys. Rev. B \ {\bf 43}, \ (1991)\ 2063.

\bibitem{Nikolaev99} P. Nikolaev, M. J. Bronikowski, R. K. Bradley, F. Rohmund,
D. T. Colbert, K.A. Smith, R. E. Smalley, Chem. Phys. Lett. {\bf
313}, (1999) 91.

\bibitem{Daniels70} J. Daniels, C. von Festenberg, H. Raether, and K. Zeppenfeld,
Optical Constants of Solids by Electron Spectroscopy, in Springer
Tracts in Modern Physics Vol. 54 Springer, Berlin, 1970, pp.
126–130.

\bibitem{Ahuja97} R. Ahuja, S. Auluck, J. M. Wills, M. Alouani, B. Johansson, O.
Eriksson, Phys. Rev. B. {\bf 55}, (1997) 4999.

\bibitem{Lin00} Lin, M. F. Phys. Rev. B {\bf 61}, (2000) 13153.

\bibitem{Hartschuh03} A.\ Hartschuh,\ H.\ N.\ Pedrosa, L.\ Novotny,\ T.\ D.\ Krauss, \ Science \ {\bf
301}, \ (2003)\  1354.

\bibitem{Gregan02} E. Gregan, S. M. Keogh, T. G. Hedderman, B. McCarthy, G. Farrell,G. Chambers, H. Byrne, J. AIP Conference
Proceedings {\bf 633}, (2002)
 294.

\bibitem{Kukovecz02} A. Kukovecz, Ch. Kramberger, V. Georgakilas, M. Prato, H. Kuzmany, Eur. Phys. J. B {\bf 28}, (2002) 223.

\bibitem{Ding02} J. W. Ding, X. H. Yan, J. X. Cao, Phys. Rev. B {\bf 66}, (2002) 073401.

\bibitem{Saito92} R. Saito, M. Fujita, G. Dresselhaus, and M. S. Dresselhaus, Phys. Rev. B {\bf 46}, (1992) 1804.

\bibitem{Hamada92} N. Hamada, S. Sawada, and A. Oshiyama, Phys. Rev. Lett. {\bf 68}, (1992) 1579.

\bibitem{Odom98} T.W. Odom, J.-L. Huang, P. Kim, and C. M. Lieber, Nature {\bf 391},
(1998) 62.

\bibitem{Wildoer98} J. W. G. Wildoer, L. C. Venema, A. G. Rinzler, R. E. Smalley,
C.  Dekker, Nature {\bf 391}, (1998) 59.

\bibitem{Liu02}  X. Liu, T. Pichler, M. Knupfer, M. S. Golden, J. Fink, H. Kataura,Y. Achiba, Phys. Rev. B {\bf 66}, (2002) 045411.

\bibitem{Dresselhaus99} G. Dresselhaus, M. A. Pimenta, R. Saito, J. C. Charlier, S. D. M. Brown,et al. In: Science and Application of Nanotubes; D. Tomanek, R. J. Enbody,  Eds.; Kluwer Academic, New York, 1999; pp
275-295.

\bibitem{Ando97} T. Ando, J. Phys. Soc. Jap. {\bf 66}, (1997)
1066.

\bibitem{Hertel96} T. Hertel, E. Knoesel, M. Wolf, and G. Ertl,
Phys. Rev. Lett. {\bf 76}, (1996) 535.

\bibitem{Moos01} G. Moos, \ C. Gahl, \ R. Fasel, \ M. Wolf and \ T. Hertel, \ Phys. Rev. Lett. \ {\bf 87}, \ (2001)\ 267402.

\bibitem{Spataru01} C. D. Spataru,M. A. Cazalilla, A. Rubio, L. X. Benedict, P. M. Echenique,and S. G. Louie, \ Phys. Rev.
Lett. \ {\bf 87}, \ (2001)\  246405.

\bibitem{Allen87} P.B. Allen, Phys. Rev. Lett. {\bf 59}, (1987)
1460.

\bibitem{Brorson90} S.D. Brorson, A. Kazeroonian, J. S. Moodera, D.W. Face, T.K. Cheng, E.P. Ippen, M.S. Dresselhaus, and G. Dresselhaus, Phys. Rev. Lett. {\bf 64}, (1990)
2172.

\bibitem{Allen75} P.B. Allen, Phys. Rev. B {\bf 11}, (1975) 2693.

\bibitem{Frank98} S. Frank, P. Poncharal, Z. L. Wang, W. A. de
Heer, Science {\bf 280}, (1998) 1744.

\bibitem{Appenzeller01} J. Appenzeller, R. Martel, P. Avouris, H.
Stahl, and B. Lengeler, Appl. Phys. Lett. {\bf 78}, (2001) 3313.


\end{thebibliography}
\end{document}